\documentclass[aps,prb,twocolumn]{revtex4} 
\usepackage{amsmath} 
\usepackage{graphicx} 
\usepackage{color}

\begin{document} 
\title{Nonlocal damping of helimagnets in one-dimensional interacting electron systems} 

\author{Kjetil M. D. Hals, Karsten Flensberg and Mark S. Rudner} 
\affiliation{ Niels Bohr International Academy and the Center for Quantum Devices, Niels Bohr Institute, University of Copenhagen, 2100 Copenhagen, Denmark} 
\begin{abstract}

We investigate the magnetization relaxation of a one-dimensional helimagnetic system coupled to interacting itinerant electrons.
The relaxation is assumed to result from the emission of plasmons, the elementary excitations of the one-dimensional interacting electron system, caused by slow changes of the magnetization profile. 
This dissipation mechanism leads to a highly nonlocal form of magnetization damping that is strongly dependent on the electron-electron interaction. Forward scattering processes lead to a spatially constant damping kernel, 
while backscattering processes produce a spatially oscillating contribution. 
Due to the nonlocal damping, the thermal fluctuations become spatially correlated over the entire system. 
We estimate the characteristic magnetization relaxation times for magnetic quantum wires and nuclear helimagnets.
\end{abstract}

\maketitle 

\section{Introduction} 
Recently, intense interest has developed in the helical magnetic ordering of one-dimensional (1D) systems of local moments coupled to itinerant electrons. 
Such systems exhibit a variety of intriguing many-body phenomena, such as spin-Peierls instabilities\cite{Pincus:SSC1971,Braunecker:PRB2010} and induced topological superconductivity,\cite{Choy:PRB2011, Kjaergaard:PRB20120, Klinovaja:PRL2013} which result from the details of magnetism, electronic structure, and electron-electron interactions.
These phenomena may be relevant for a wide variety of physical systems, ranging from magnetic atoms on superconducting~\cite{Choy:PRB2011, Martin:PRB2012, Nadj-Perge:PRB2013, Vazifeh:PRL2013, Simon:PRL2013, Nadj-Perge:Science2014} and normal metal~\cite{Menzel:PRL2012} substrates, to single walled carbon nanotubes~\cite{Braunecker:PRL2009} and semiconductor-based quantum wires.~\cite{Braunecker:PRB2009, Klinovaja:PRL2013, Scheller:PRL2014}

While much of the work in this area so far has focused on static and thermodynamic properties of the 1D helimagnets, a richer understanding may be gained by developing and employing new {\it dynamical} probes for assessing the behaviors of these systems. For example, an interesting self-tuning effect was proposed for systems dominated by a Ruderman-Kittel-Kasuya-Yoshida (RKKY)-type interaction:~\cite{Braunecker:PRL2009, Braunecker:PRB2009, Braunecker:PRB2012, Scheller:PRL2014} the local moments are predicted to order into a spiral arrangement which, through coherent backscattering, gaps out one spin channel of the itinerant electron system for {\it any value} of the electron density.
This remarkable phenomenon was even suggested as providing a route towards realizing topologically protected Majorana bound states in quantum wires.~\cite{Vazifeh:PRL2013, Klinovaja:PRL2013, Simon:PRL2013}
However, because direct probes of magnetization are unavailable for many systems, it can be challenging to positively identify this intriguing magnetic state (necessarily via indirect means).~\cite{Scheller:PRL2014} With further theoretical understanding of dynamical responses, such as typical damping or relaxation times, additional tests (e.g., density quenches which change the preferred ordering wave vector) could be used to clarify the natures of the underlying states. 

More generally, magnetization relaxation processes determine the magnetic response to external perturbations as well as to spontaneous thermal fluctuations.
Furthermore, the nature of the magnetic response is crucially important for noise and magnetization dynamics in magnetoelectronic devices.~\cite{Bland:book, Reviews:STT} 
A better understanding of the spin dynamics in 1D helimagnets may pave the way for exploring phenomena such as current-driven magnetization dynamics, with potential practical applications beyond those envisaged so far. 
Thus, the investigation of microscopic damping mechanisms is essential for developing a thorough fundamental and practical understanding of these exciting new magnetic systems. 

\begin{figure}[t] 
\centering 
\includegraphics[scale=1.0]{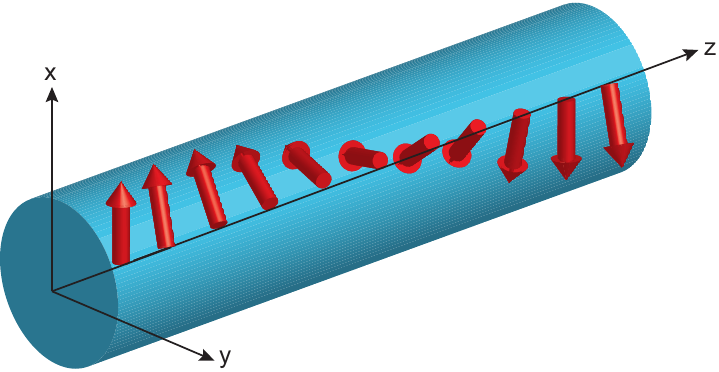}  
\caption{(Color online). Helimagnet formed in a 1D conductor.}
\label{Fig1} 
\end{figure} 
Given the motivations above, in this work we investigate the relaxation of 1D helimagnets via the emission of collective 
excitations into the interacting itinerant electron system. 
Note that the 1D nature of the itinerant system is important -- our theory is meant to describe quasi-1D systems with a single transverse mode at the Fermi energy (e.g., semiconductor quantum wires~\cite{Scheller:PRL2014}).
The elementary excitations of these 1D electronic systems are plasmons, which describe density waves.  
Interestingly, previous theoretical works have predicted that electron-electron interactions in such 1D systems may play important roles both in establishing ordering~\cite{Braunecker:PRL2009, Braunecker:PRB2009} and in the relaxation dynamics of weakly-coupled (non-ordered) nuclear spins.~\cite{Kiss:PRL2011, Zyuzin:arxiv}
In this work, we use a bosonization approach to study the non-perturbative effects of electron-electron interactions on the damping of {\it ordered} spins. We find that interactions have a profound effect on damping, leading to an
enhancement of the damping by several orders of magnitude. The damping has a highly non-local character. Consequently, the thermal fluctuations become spatially correlated over the entire sample.
We estimate the characteristic magnetization relaxation times due to this mechanism for two classes of systems:   (Ga,Mn)As quantum wires and nuclear helimagnets formed in GaAs quantum wires.

\section{Theory and model}
Our approach is based on the theoretical framework developed for magnetization damping in metallic ferromagnets.~\cite{Brataas:PRL2008, Brataas:PRB2011} 
A key ingredient of the model is that the dynamics of the low-lying collective spin excitations 
are parametrized by a classical magnetization order-parameter field whose magnitude is assumed to be constant in time and homogeneous in space, while its local orientation is allowed to fluctuate. In this case, the evolution of the spin system can be described by the Landau-Lifshitz-Gilbert (LLG) phenomenology:~\cite{Bland:book, Reviews:STT, comment:CoarseGrain} 
\begin{eqnarray}
\dot{\mathbf{m}} (z,t) &=& -\gamma \mathbf{m} (z,t) \times [ \mathbf{H}_{\rm eff} (z, t) + \mathbf{h}_T (z,t)  ]  + \nonumber \\
 & & \mathbf{m} (z,t)\times \int { dz' } \boldsymbol{\alpha} (z, z') \dot{\mathbf{m}} (z', t). \label{Eq:LLG}
\end{eqnarray}
Here, the unit vector $\mathbf{m} (z, t)$ parametrizes the local spin-order and is oriented parallel to the magnetization vector $\mathbf{M} (z, t)= M_s\mathbf{m} (z, t)$, $\gamma= g\mu_B/\hbar$ is the gyromagnetic ratio in terms of the g-factor of local spins and the Bohr magneton $\mu_B$, and $\mathbf{H}_{\rm eff}= -\delta F/\delta \mathbf{M}$ is the effective field found by varying the magnetic free energy functional $F[\mathbf{M}]$ with respect to the magnetization.
The quantity $ \mathbf{h}_T (z,t)$ in the first line is a stochastic magnetic field induced by the thermal fluctuations (to be discussed further below). 
We assume that the free energy functional stabilizes an equilibrium helimagnetic texture of the form~\cite{Braunecker:PRL2009, Braunecker:PRB2009, Vazifeh:PRL2013, Klinovaja:PRL2013, Menzel:PRL2012}
\begin{equation}
\label{Eq:m0} \mathbf{m}_0 (z)= [\cos (qz), \sin (qz), 0] ,
\end{equation}
where $q$ depends on the ordering mechanism.
Throughout this work, we use coordinate axes with the $z$-axis oriented along the 1D conductor (see Fig.~\ref{Fig1}).

Magnetization relaxation is described by the second-rank Gilbert damping tensor $\alpha_{ij} (z, z')$ in Eq.~\eqref{Eq:LLG}. 
We consider magnetization relaxation via excitations of the itinerant electron system.
In this case, the Gilbert damping tensor is given by~\cite{Brataas:PRL2008, Brataas:PRB2011} (see Appendix~\ref{app1} for a derivation) 
\begin{eqnarray}
\alpha_{ij} (z, z')=  -\frac{4\gamma h_0^2}{\hbar^2 M_s}  \lim_{\omega\rightarrow 0}  \frac{\Im m\left[ \chi_{ij} ( z, z', \omega) \right] }{\omega},  \label{Eq:alphaExpr}
\end{eqnarray}
where $\chi_{ij} (z, z',\omega)= \int_{-\infty}^{\infty}{ dt}\, \chi_{ij} ( z, z', t) \exp(i\omega t)$ is the Fourier transform of the spin susceptibility of the itinerant electrons, 
$\chi_{ij} (z, z',t)= -(i/\hbar) \theta(t) [\hat{s}_i (z,t), \hat{s}_j (z',0)]$. 
Here, $\hat{\mathbf{s}} (z,t) =  (\hbar/2) \boldsymbol{\psi}^{\dagger} (z, t) \boldsymbol{\sigma}  \boldsymbol{\psi}(z, t)$ is the spin-density operator for itinerant electrons, taken in the interaction picture with respect to Hamiltonian (\ref{Eq:Hamiltonian}) below, with the {\it static} magnetization \eqref{Eq:m0}. Above, $\boldsymbol{\sigma}$ is the vector of Pauli matrices and $\boldsymbol{\psi}(z)\!=\![\psi_{\uparrow} (z), \psi_{\downarrow} (z)]$ is the spinor-valued fermionic field operator.

We model the itinerant electrons via the Hamiltonian:
\begin{eqnarray}
H &=& \int { dz}\, \boldsymbol{\psi}^{\dagger} (z) \left[ \frac{\hat{p}_z^2}{2m} + h_0\mathbf{m}(z,t)\cdot\boldsymbol{\sigma}  \right] \boldsymbol{\psi}(z) +  \label{Eq:Hamiltonian} \\
&&\!\!\!\!\!\!\!\! \frac{1}{2} \iint { dz dz'}\, \psi_{\sigma}^{\dagger} (z) \psi_{\sigma'}^{\dagger} (z') V_{\rm ee} (z-z')\psi_{\sigma'}(z')\psi_{\sigma}(z) ,\nonumber 
\end{eqnarray}
where $\hat{p}_z$ is the momentum operator, $V_{\rm ee}$ is the electron-electron interaction potential, and $h_0$ is the magnetic coupling.
Summation over repeated indices is implied. 

In the calculation below, we aim to evaluate the Gilbert damping tensor in Eq.~\eqref{Eq:alphaExpr}, using the spin susceptibility for the electronic system described by Eq.~\eqref{Eq:Hamiltonian}, with a fixed chemical potential.  Linearizing around the Fermi points, we will develop a Luttinger liquid type description of the nearly helical system, allowing interactions to be taken into account non-perturbatively.

\section{Results}
\begin{figure}[t] 
\centering 
\includegraphics[scale=1.0]{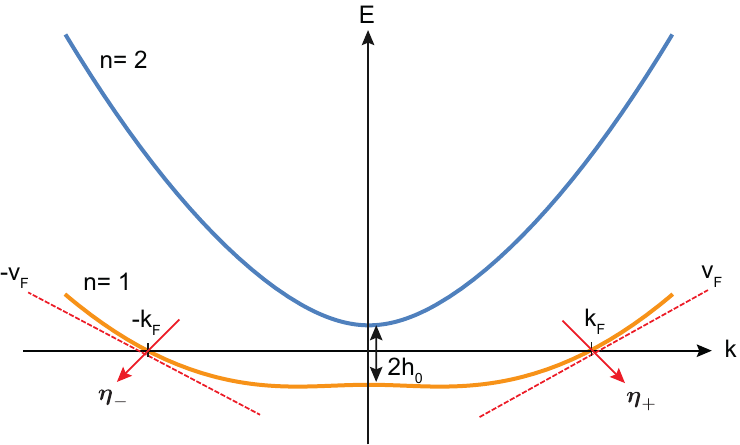}  
\caption{(Color online). Energy dispersion of the gauge transformed free-electron Hamiltonian $H_u^{(0)}$. The bosonization is performed by linearizing the dispersion about the Fermi wavevectors $k= \pm k_F$ and fixing  
the k-dependent eigenspinors $\boldsymbol{\eta}_r$ to their values at the Fermi energy. }
\label{Fig2} 
\end{figure}

We now explicitly calculate the Gilbert damping tensor in Eq.~\eqref{Eq:alphaExpr}.
To facilitate calculation of the spin susceptibility, we transform to a non-uniformly rotated frame via the unitary transformation $\boldsymbol{\psi}_u= U(z) \boldsymbol{\psi}$, with $U(z) = e^{i q z \sigma_z/2}$.
This transformation ``untwists'' the helix, rendering the free electron part of the transformed Hamiltonian $H_u= UHU^{\dagger}$ translationally invariant,
\begin{equation}
H_u^{(0)} = \int { dz}\, \boldsymbol{\psi}_u^{\dagger}(z)\! \left[ \frac{\hat{p}_z^2}{2m} - \frac{\hbar q}{2m}\sigma_z \hat{p}_z +  h_0 \sigma_x \right]\! \boldsymbol{\psi}_u(z),  \label{Eq:Hamiltonian_U} 
\end{equation}
while the interaction term is unaffected.
In this representation, the spin susceptibility and the Gilbert damping tensor transform to $\boldsymbol{\chi}_u (z,z',t)= R(z) \boldsymbol{\chi} (z,z',t)R^T(z')$ and $\boldsymbol{\alpha}_u (z,z',t)= R(z) \boldsymbol{\alpha} (z,z',t)R^T(z')$, where $R(z)$ is the SO(3) matrix associated with $U(z)$.
The energy dispersion of $H_u^{(0)}$ is shown in Fig.~\ref{Fig2}; its eigenfunctions are $\boldsymbol{\psi}_{n, k} (z)= \boldsymbol{\eta}_{n, k}\otimes \psi_{n,k} (z)$, where $n\in \left\{ 1,2\right\}$ is the band index, $ \boldsymbol{\eta}_{n, k}$ is the eigenspinor,  and $\psi_{n,k} (z)= \exp (ikz)/\sqrt{L}$, for a system of length $L$.

In this work, we set the chemical potential in the gap that separates the bands near $k = 0$, 
such that the single-particle dispersion in Eq.~\eqref{Eq:Hamiltonian_U} features only a {\it single branch} of right and left moving modes at the Fermi energy. 
We neglect interband couplings and write an effective description within the lowest band, linearized 
about the Fermi wavevectors $\pm k_F$. We fix the spinor parts of the wave functions to their values at the Fermi energy (Fig.~\ref{Fig2}).~\cite{Comment:OneBand}

To compute the spin-spin susceptibility in the presence of electron-electron interactions, we employ a bosonic description. 
As a first step, we express the fermionic field operator (projected into the lowest band) as a superposition of fields representing right (+) and left (-) movers:  $\boldsymbol{\psi}_u (z) = \boldsymbol{\psi}_{+} (z) + \boldsymbol{\psi}_{-} (z)$. 
The fields $\boldsymbol{\psi}_{r}$ ($r\in \left\{ +, - \right\} $) take the form $\boldsymbol{\psi}_{r} = \boldsymbol{\eta}_{r}\otimes \psi_{r} (z)$, where $\boldsymbol{\eta}_{r}= \boldsymbol{\eta}_{1, r k_F}$  and the spatial part (in terms of the destruction operators $c_{k,r}$) is 
$\psi_{r} (z)=\frac{1}{\sqrt{L}} e^{irk_F z}\,\sum_{k} e^{ikz} c_{k,r}$. 
Substituting the fermionic field operator into the Hamiltonian $H_u$, performing a Fourier transformation to k-space, and evaluating $V_{ee} (q)$ at momentum zero and $2k_F$ for forward- and back- scattering processes, respectively, we obtain     
\begin{equation}
H_u = \sum_{k,r}r\hbar v_F k c^{\dagger}_{k,r} c_{k,r} + \sum_{q,r}( g_2 \rho_{q,r}\rho_{-q,-r} + g_4 \rho_{q,r}\rho_{-q,r}).  \label{Eq:Hamiltonian_U2} 
\end{equation}
Here,  $v_F= \hbar k_F/m - \hbar q \boldsymbol{\eta}^{\dagger}_{+}\sigma_z \boldsymbol{\eta}_{+}/2m$, $g_2= (V_{ee}(0) -  |\boldsymbol{\eta}^{\dagger}_{+} \boldsymbol{\eta}_{-}|^2 V_{ee} (2k_F))/2L$, $g_4= V_{ee} (0)/2L$, and $\rho_{q,r}= \sum_{k} c^{\dagger}_{k-q,r} c_{k,r}$ is the Fourier-transformed density operator for right/left-movers.    
Following the standard procedure,~\cite{Giamarchi:book} we write Eq.~\eqref{Eq:Hamiltonian_U2} in the bosonized form
\begin{eqnarray}
H_u &=& \frac{\hbar  }{2\pi} \int { dz}\, \left[ u_{\rm eff} K (\partial_{z} \theta)^2  + \frac{u_{\rm eff}}{K}(\partial_{z} \phi)^2   \right] ,  \label{Eq:HBosonized}
\end{eqnarray} 
where $\phi$ and $\theta$ are the bosonic fields, $u_{\rm eff}$ is the density wave velocity, and $K$ is the Luttinger parameter. 

The bosonic representations of the fermionic fields are
\begin{eqnarray}
\boldsymbol{\psi}_{r} (z)= \boldsymbol{\eta}_{r}\otimes \frac{U_{r}}{\sqrt{2\pi a}} e^{i r k_F z } e^{- i [ r\phi (z) - \theta (z)  ]},  \label{Eq:Bosonization}
\end{eqnarray}
where $a$ is an infinitesimal short distance cutoff~\cite{comment1} and $\{U_{r}\}$ are the Klein factors.
The repulsive electron-electron interaction implies that $0 < K \leq 1$, where $K = 1$ for non-interacting electrons.

We calculate the spin susceptibility following the standard approach for Luttinger liquids, see Ref.~\onlinecite{Giamarchi:book} for details. 
The resulting (imaginary) time-ordered spin-spin correlation function~\cite{commentM1}, $\tilde{\chi}_{u,ij} (z, z',\tau) = - \langle T_{\tau} \hat{s}_i (z,\tau) \hat{s}_j (z',0)  \rangle$, 
is diagonal and can be written as $\tilde{\chi}_{u,ii} (\tilde{z},\tau) = \tilde{\chi}^{(0)}_{u,ii} (\tilde{z},\tau) + \tilde{\chi}_{u,ii}^{(2k_F)} (\tilde{z},\tau) \cos (2k_F \tilde{z} ),$ where 
\begin{eqnarray}
\tilde{\chi}_{u,ii}^{(0)} &=& - \left(\frac{\hbar}{2\pi}\right)^2 \frac{K^{(i)} \Lambda_{ii}^{++}}{2} \frac{\upsilon^2 - \tilde{z}^2 }{(\tilde{z}^2 + \upsilon^2)^2}, \\ 
\label{Eq:chiii}\tilde{\chi}_{u,ii}^{(2k_F)} &=& - \left(\frac{\hbar}{2\pi}\right)^2  \frac{\Lambda_{ii}^{+-}}{2a^2}  \left( \frac{a^2}{\tilde{z}^2 + \upsilon^2}\right)^K .
\end{eqnarray} 
Here, $\tilde{z}= z - z'$, $\upsilon= u_{\rm eff} \tau + a\, {\rm sign}(\tau)$, $K^{(x)}=K^{(y)}=K$, $K^{(z)}= K^{-1}$, and  $\Lambda_{ii}^{r r'} = |\boldsymbol{\eta}^{\dagger}_{r}\sigma_i \boldsymbol{\eta}_{r'}|^2$.
The spin susceptibility is $\chi_{u,ii} (\tilde{z},t) = -(2/\hbar)\theta(t) \Im {\rm m} [\tilde{\chi}_{u,ii} (\tilde{z}, t)]$, where $\tilde{\chi}_{u,ii} (\tilde{z}, t)$
is the time-ordered correlation function in {\it real time}, which is obtained via the Wick rotation $\tau= it + 0^{+} {\rm sign} (t)$. 
For $K<1$,  $\chi_{u,ii}^{(2k_F)}(\omega)/\omega$ diverges in the low-frequency limit. 
We regularize the divergence by evaluating the expression at the low-frequency cut-off $\omega_{0}= u_{\rm eff}2 \pi/L$ set by the finite length  of the system.~\cite{comment2b}

The analysis above gives the Gilbert damping tensor
\begin{eqnarray}
\alpha_{u,ii} (z,z') &=& \alpha^{(0)}_{u,ii} + \alpha^{(2k_F)}_{u,ii} \cos (2k_F\tilde{z}) , \label{Eq:alpha} \\
\alpha^{(0)}_{u,ii}  &=& \frac{ \gamma h_0^2 K^{(i)} \Lambda_{ii}^{++} }{2\pi\hbar M_s u_{\rm eff}^2 }  ,  \label{Eq:alpha_1} \\
\alpha^{(2k_F)}_{u,ii} &=&\frac{\gamma h_0^2 \Lambda_{ii}^{+-}  F_{K} \left( \zeta   \right) }{ 2^{1/2+K} \pi^{3/2} \hbar M_s u_{\rm eff}^{2}  \Gamma (K) } .  \label{Eq:alpha_2}
\end{eqnarray}   
Here,  $\Gamma(K)$ is the gamma function,  $\zeta\equiv a \omega_0 / u_{\rm eff}$, and  $F_{K} (\zeta) = \pi \zeta^{K - 3/2} [ (I_{K-1/2} (\zeta) - L_{1/2-K}(\zeta)) - 2\zeta \cos (K\pi) \mathcal{K}_{K-1/2} (\zeta) ] $, where $I_{\nu} (\zeta)$ is the modified Bessel function of the first kind, $L_{\nu}(\zeta)$ is the modified Struve function, and $\mathcal{K}_{\nu} (\zeta)$ is the modified Bessel function of the second kind.  
To obtain the lab-frame damping tensor, the transformation 
$\boldsymbol{\alpha} (z,z',t)= R^T (z) \boldsymbol{\alpha}_u (z,z',t)R(z')$ must be applied. 
We continue the analysis in the {\it rotated frame}, where the expressions are much simpler.

\section{Discussion}
Equations~\eqref{Eq:alpha} - \eqref{Eq:alpha_2} are the central results of this work, and describe magnetization relaxation of 1D helimagnets via plasmon excitations.

The damping consists of two parts with very distinct position dependencies: a homogeneous
term $\sim\alpha^{(0)}$ and a rapidly oscillating term $\sim\alpha^{(2k_F)}$. The corresponding highly nonlocal magnetization relaxation caused by the plasmon 
excitations differs markedly from the damping of conventional metallic ferromagnets, which is believed to be local.~\cite{CommentM2}

Constraints of the model reduce the number of independent tensor elements.
First, $\Lambda_{yy}^{++}= \Lambda_{zz}^{+-}= 0$ implies that the damping tensor is described by four independent coefficients:
$\alpha^{(0)}_{u,xx}$, $\alpha^{(0)}_{u,zz}$, $\alpha^{(2k_F)}_{u,xx}$, and $\alpha^{(2k_F)}_{u,yy}$. 
Second, the constraint $\dot{\mathbf{m}}\cdot\mathbf{m} =0$ imposed by normalization implies
that $\dot{m}_x=0$ in the rotated reference frame (for small $\delta\mathbf{m}$). Thus, the damping is governed by only two coefficients: $\alpha^{(0)}_{u,zz}$ and $\alpha^{(2k_F)}_{u,yy}$.  

What are the characteristics of these two independent damping coefficients?
The tensor element $\alpha^{(0)}_{zz}$ originates from forward scattering processes and governs the damping of the long wave-length spin-wave modes. 
Similarly, $\alpha^{(2k_F)}_{yy}$ is associated with electronic backscattering, and controls the relaxation of short wave-length modes.

\begin{figure}[t] 
\centering 
\includegraphics[scale=1.0]{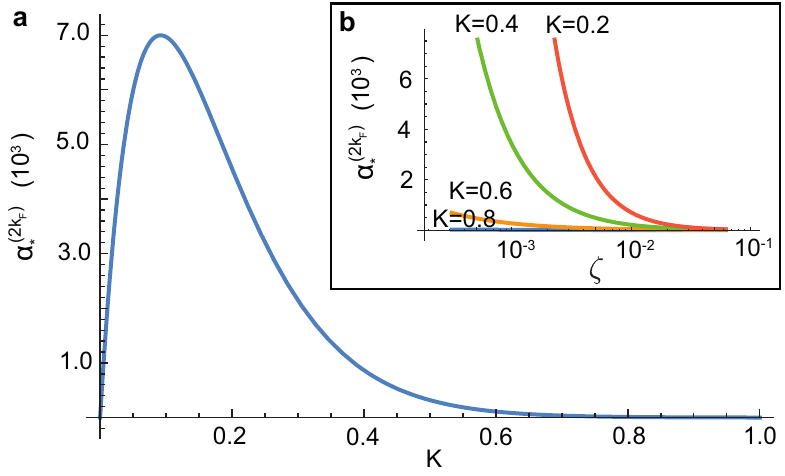}  
\caption{(Color online). (a) The dimensionless damping parameter $\alpha^{(2k_F)}_{\ast}(K,\zeta)= \alpha^{(2k_F)}_{u,yy}(K, \zeta)/ \alpha^{(2k_F)}_{u,yy}(1,\zeta)$ as a function of the electron-electron interaction parameter $K$ with $\zeta=a \omega_0 / u_{\rm eff}$ fixed at $\zeta= 3.1\times10^{-3}$. (b) The damping parameter as a function of $\zeta$ for different K.}
\label{Fig3} 
\end{figure}

Interestingly, $\alpha^{(0)}_{zz}$  is proportional to the momentum-momentum correlator $\langle \partial_{\tilde{z}}\theta (\tilde{z},\tau) \partial_{\tilde{z}} \theta (0,0) \rangle$, and is thus inversely proportional to the Luttinger parameter $K$.  Consequently, electron-electron interaction enhances  $\alpha^{(0)}_{u, zz}$ as $K^{-1}$. 
A much more complex dependency of the electron-electron interactions is seen in the magnetization damping caused by backscattering processes. 
Remarkably, electron-electron interactions may increase $\alpha^{(2k_F)}_{yy}$ by nearly four orders of magnitude compared to its value in the non-interacting limit $K=1$ (Fig.~\ref{Fig3}a).  
The tensor element $\alpha^{(2k_F)}_{yy}$ reaches a maximum at  $K\approx 0.1$ before it drops quickly to zero in the strongly interacting regime $K\rightarrow 0$.  
In this limit, the potential energy $(u_{\rm eff}/K)(\partial_z \phi )^2$ of the bosonic Hamiltonian \eqref{Eq:HBosonized} completely governs the electron dynamics and density variations of the Luttinger liquid become insusceptible to time variations in the magnetization. The dramatic enhancement of $\alpha^{(2k_F)}_{yy}$ is related to the fact that electron-electron interactions make the damping extremely sensitive to $\zeta\sim a/L$, which measures the ratio of the short distance cut-off  to the long distance cut-off  (Fig.~\ref{Fig3}b). 
In the absence of interactions, $\alpha^{(2k_F)}_{yy}$ approaches a constant value in the limit $\zeta \rightarrow 0$. However, with interactions,  $\alpha^{(2k_F)}_{yy}$ is singular in this limit and the singularity becomes stronger with increasing strength of the interactions. 

To investigate the experimental consequences of Eqs.~\eqref{Eq:alpha}-\eqref{Eq:alpha_2},
we discuss thermal fluctuations and estimate the characteristic relaxation time for two classes of systems which are proposed to hold 1D helimagnetic states.
Additionally, we propose a method for probing the relaxation time in nuclear wires via transport measurements.  
The magnetic order is assumed to be to stabilized by the RKKY interaction, implying 
$q= k_F$.~\cite{Comment_kf}   

The form of the damping tensor has remarkable implications for the statistical properties of thermal fluctuations.
While the average of the stochastic magnetic field $\mathbf{h}_T$ in Eq.~\eqref{Eq:LLG} is zero, $\langle \mathbf{h}_T \rangle = 0$, its correlations (in accordance with the fluctuation-dissipation theorem) are given in the classical (Maxwell-Boltzmann) limit as~\cite{Brown:PR1963, Brataas:PRB2011} 
\begin{eqnarray}
\langle h_{T, i} (z,t) h_{T, j} (z',t') \rangle = \frac{2 k_B T}{\gamma M_s} \alpha_{ij} (z, z') \delta (t-t') , \label{Eq:hCorr}
\end{eqnarray}
where $T$ is the temperature and the average $\langle ... \rangle$ is taken over an ensemble in thermal equilibrium. 
According to Eq.~\eqref{Eq:hCorr}, the thermal fluctuations are correlated over the entire sample. The fluctuations divide into two distinct classes; one class characterized by a spatially constant correlation and a second class characterized by an oscillating $\sim\cos (2k_F\tilde{z})$ correlation. Strongly correlated thermal fluctuations of this form have not been reported or investigated before in any magnetic system, and  
a thorough investigation of how the associated stochastic field $\mathbf{h}_T (z)$ in Eq.~\eqref{Eq:LLG} influences the magnetization dynamics should be an interesting task for future studies.

For a magnetization precessing at frequency $\omega$, Eq.~\eqref{Eq:LLG} yields two characteristic relaxation times $\tau_{\rm rel}^{(0)}$ and $\tau_{\rm rel}^{(2k_F)}$, associated with the homogeneous and oscillatory dissipation terms:  
$\tau_{\rm rel}^{(0, 2k_F)} \sim [ \alpha^{(0,2k_F)} L \omega ]^{-1}$. 
We now estimate $\tau_{\rm rel}^{(0, 2k_F)}$ for two classes of systems, which are believed to host 1D helimagnetism:~\cite{Klinovaja:PRL2013} 
 Ga$_{0.98}$Mn$_{0.02}$As quantum wire and GaAs wire in which the nuclear spins are hyperfine coupled 
to the itinerant system (see Appendix~\ref{appParaValues} for material parameters). The (Ga,Mn)As and GaAs wire cross sections are assumed to contain $50\times50$ unit cells.
The characteristic magnon frequency of the first excited mode is $\omega= k_B T_c/ \hbar I_{\bot}^{1/(3-2K)}$,~\cite{Braunecker:PRB2009} 
where $T_c$ is the critical temperature and $\hbar I_{\bot}$ is the total spin of the cross section. 
We assume that each unit cell is fully spin polarized.
For the magnetic (Ga,Mn)As wire, we find $\tau_{\rm rel}^{(0)} = 2.3\times 10^{-8}$~s and  $\tau_{\rm rel}^{(2k_F)}= 7.6\times 10^{-11}$~s, while for the nuclear wire
the relaxation times are $\tau_{\rm rel}^{(0)} = 1.0$~s and  $\tau_{\rm rel}^{(2k_F)}= 6.4\times 10^{-3}$~s.  

Dynamical probes offer new routes for characterizing the nature of 1D helimagnetic systems, providing complementary information to that obtained in static (dc) measurements.
For example, consider the recent experiment of Ref.~\onlinecite{Scheller:PRL2014}, in which the conductance of a GaAs-based quantum wire was observed to drop 
from $2e^2/h$ to $e^2/h$ when the temperature was reduced to below 0.1 K.
The appearance of an $e^2/h$ plateau hints at the lifting of spin-degeneracy, and was interpreted as evidence of the formation of a nuclear spin helix.
 Importantly, within the model used to interpret the experiment, the spatial period of the helix tunes itself to be equal to half of the Fermi wavelength, i.e., it is directly linked to the density  
of the electronic system.
Furthermore, the appearance of an $e^2/h$ plateau is directly linked to this commensuration between the ordering and Fermi wavevectors.
A rapid change of backgate voltage which alters the electron density should thus destroy the commensurability; just after such a quench, one would then expect to observe a conductance of $2e^2/h$, which would gradually return to a value of $e^2/h$ as the nuclear system finds its new equilibrium order, on a timescale set by the magnetic relaxation rate.
Taking similar assumptions and parameter values to those of the model used to support the conclusions of Ref.~\onlinecite{Scheller:PRL2014}, we predict very long relaxation times for the nuclear magnetic order, on the order of milliseconds or more.
Nuclear spin diffusion, another possible important relaxation mechanism, is expected to become effective on longer timescales (see Appendix~\ref{app2}).
Dynamics on such timescales should in principle be observable in experiments, and therefore may provide crucial independent means for verifying the interpretation of the experiment.

In conclusion, we have developed a theoretical formalism for describing magnetization dissipation of 1D helimagnets via the emission of plasmon excitations. 
The damping is found to be highly nonlocal and strongly dependent on the electron-electron interaction, differing markedly from the damping of conventional metallic ferromagnets. 

\section{Acknowledgements.}
Research supported the Danish National Research Foundation. 
MR acknowledges support from the Villum Foundation and the People Programme (Marie Curie Actions) of the European Union's Seventh Framework Programme (FP7/2007-2013) under REA grant agreement PIIF-GA-2013-627838.

\appendix

\section{Derivation of damping tensor}\label{app1}
We calculate the magnetization-damping rate by relating it to the energy absorption rate of the itinerant electron system subjected to small fluctuations about a static helimagnetic profile $\mathbf{m}_0(z)$.
In describing the interaction between the magnetic order parameter field and the itinerant electron system, we assume that the magnetization evolves very little over the characteristic timescales of electron dynamics.
In this case, the response of the electron system can be calculated using linear response (Kubo) theory.  

The total energy dissipation of the magnetic system is 
\begin{equation}
\dot{E}= \int dz\, \dot{\mathbf{M}} \cdot \frac{\delta F}{ \delta \mathbf{M}}. 
\end{equation}
The time derivative $\dot{\mathbf{M}} $ is determined by the LLG equation, which yields the energy dissipation 
\begin{eqnarray}
\dot{E} (t)= -\frac{M_s}{\gamma} \iint { dz dz'}\, \dot{\mathbf{m}} (z,t)\cdot [\tilde{\boldsymbol{\alpha}} (z, z') \dot{\mathbf{m}} (z', t) ] . \label{Eq:Edot}
\end{eqnarray}

Due to energy conservation, the energy lost by the magnetic system must be gained by the itinerant electron system to which it is coupled.
This implies that $\dot{E} = - \langle \dot{H} (t) \rangle$, where $H(t)$ is the Hamiltonian \eqref{Eq:Hamiltonian} of the itinerant system coupled to the  magnetization 
$\mathbf{m}(z,t) = \mathbf{m}_0(z) + \delta\mathbf{m}(z,t)$. 

We now use linear response theory to obtain the rate of change of the electronic energy due to a slow evolution of the magnetization $\mathbf{m}(z,t)$:~\cite{Rammer:book}    
\begin{eqnarray}
 \langle \dot{H} (t) \rangle= -\frac{i}{\hbar} \int_{-\infty}^{\infty} {dt'} \theta(t-t') \langle [ \dot{H} (t), \delta H (t')] \rangle, \label{Eq:Hdot1}
\end{eqnarray}
where $\theta(t)$ is the Heaviside step function, and  $\delta H (t')= (2h_0/\hbar) \int {dz'}  \delta\mathbf{m}(z',t')\cdot\hat{\mathbf{s}} (z',t')$ is the perturbing Hamiltonian produced by the small variation $\delta\mathbf{m}(z',t')$  of the helimagnetic order. 
Here, $\hat{\mathbf{s}} (z',t')$ is the spin-density operator $\hat{\mathbf{s}} (z) = (\hbar/2) \boldsymbol{\psi}^{\dagger} (z) \boldsymbol{\sigma}  \boldsymbol{\psi}(z)$, taken
in the interaction picture with respect to the unperturbed Hamiltonian. 

Fourier transforming Eq.~\eqref{Eq:Hdot1} with respect to time 
and using $\dot{H} (t)=  (2h_0/\hbar) \int { dz}\,  \dot{\mathbf{m}}(z,t)\cdot\hat{\mathbf{s}} (z,t)$ yields
\begin{widetext}
\begin{equation} 
-i\omega  \langle H (\omega) \rangle = \frac{1}{2\pi} \left( \frac{2h_0}{ \hbar} \right)^2  \iint { dz dz'} \int {d\omega'} i(\omega - \omega') m_i (z, \omega - \omega') \frac{i \chi_{ij}(z, z', \omega')}{\omega' } i\omega'\delta m_j (z', \omega'),
\end{equation}
\end{widetext}
where $\chi_{ij} (z, z',\omega)= \int_{-\infty}^{\infty}{ dt}\, \chi_{ij} ( z, z', t) \exp(i\omega t)$ is the Fourier transform of the spin susceptibility 
$\chi_{ij} (z, z',t)= -(i/\hbar) \theta(t) [\hat{s}_i (z,t), \hat{s}_j (z',0)]$. 
To leading order in the precession frequency, the behavior of the energy change is captured by replacing $i \chi_{ij}(\omega)/\omega$ by its value in the zero frequency limit. 
Transforming back to the time domain and using $\delta\dot{m}_j = \dot{m}_j$ gives the energy absorption rate
\begin{eqnarray}
 \langle \dot{H} (t) \rangle= \frac{4 h_0^2}{\hbar^2} \iint { dz dz'} \dot{m}_i \left[\lim_{\omega\rightarrow 0} i\frac{ \chi_{ij} (z, z', \omega)}{\omega}   \right] \dot{m}_j .  \label{Eq:Hdot2}
\end{eqnarray}
Comparing Eq.~\eqref{Eq:Hdot2} with Eq.~\eqref{Eq:Edot}, we identify the following expression for the Gilbert damping tensor: 
\begin{eqnarray}
\alpha_{ij} (z, z')=  -\frac{4\gamma h_0^2}{\hbar^2 M_s}  \lim_{\omega\rightarrow 0}  \frac{\Im m\left[ \chi_{ij} ( z, z', \omega) \right] }{\omega}.  
\end{eqnarray}

\section{Material Parameters }\label{appParaValues}
Table \ref{tab:1} shows typical material parameters of (Ga,Mn)As quantum wires and nuclear helimagnets formed in GaAs quantum wires. These parameter values are used in the main text  
to estimate the characteristic relaxation times. 
Here, $L$ is the length of the quantum wire, $\mu_F$ is the chemical potential, $a$ is the short distance cutoff,~\cite{comment1}
$K$ is the Luttinger parameter, $u_{\rm eff}$ is the density wave velocity,  $M_s/\gamma$ is the spin density of the 1D magnetic system,
$h_0$ describes the coupling between the itinerant electrons and the ordered spin system, $I_{\bot}$ is the total spin contained in the cross section of the wire, and $T_c$ is the
critical temperature of the spin system.~\cite{Braunecker:PRB2009}  

The short distance cutoff (given by the chemical potential) together with the length of the wire and the density wave velocity determine 
the value of the dimensionless $\zeta$ parameter, $\zeta= a \omega_0 / u_{\rm eff}$. Because the damping becomes highly sensitive to the values of $\zeta$ and $K$ in the presence of interactions (Fig.~\ref{Fig3}), we believe uncertainties in these two parameters govern the sensitivity in the relaxation times estimated from the values in Table \ref{tab:1}.  While $\alpha^{(0)}_{\rm rel}$ only depends on $K$, the backscattering term $\alpha^{(2k_F)}_{\rm rel}$ grows as $\zeta$ becomes small (see Fig.~\ref{Fig3}), rising sharply for smaller $K$ (i.e., stronger interactions).  For fixed $\zeta$, the relaxation is strongly enhanced by interactions up to a max value and then falls off sharply. 

\begin{table}
\centering
\caption{Estimates of material parameters for two classes of systems (adapted from Refs.~\onlinecite{Klinovaja:PRL2013} and \onlinecite{Braunecker:PRB2009} ). }
\begin{tabular}{ l c c }
\hline\hline \\
  &  \ \ \ \ \  Magnetic wire \ \ \ \ \   & \ \ \ \ \  Nuclear wire \ \ \ \ \    \\
 \hline \\
 Material & (Ga,Mn)As & GaAs \\ 
L~($\mu$m) & 20 & 20 \\ 
 $\mu_F$~(meV) & 20 & 5.6 \\ 
 $a$~(nm) & 5 & 10 \\
 $K$ & 0.5 & 0.5 \\
 $u_{\rm eff}$~(ms$^{-1}$) & $3.2\times 10^{5}$ & $1.7\times 10^{5}$ \\
 $M_s/\gamma$~(Jsm$^{-1}$) & $2.3\times 10^{-23}$ & $2.8\times 10^{-21}$ \\
 $h_0$~(meV) & 5 & 0.07 \\
  $I_{\bot}$ & 500 & 15000 \\
 $T_c$~(K) & 2 & 0.01 \\	
 \hline\hline \\ 
\end{tabular}
\label{tab:1}
\end{table}

\section{Relaxation via Nuclear Spin Diffusion}\label{app2}
Let $\tau^{(d)} $ denote the characteristic magnetization relaxation time induced by nuclear spin diffusion.
The spin diffusion constant $D$ of GaAs has been estimated to be on the order of $D\sim 10^{-17}$ m$^{2}$/s.~\cite{Paget:PRB1982}
The relaxation time $\tau^{(d)} $ is the time required to diffuse from an internal point of the quantum wire to the surrounding nuclei, i.e., a diffusion length of about the wire diameter $d\sim 50 a_0$ (where $a_0$ is the lattice constant of GaAs).
This leads to following estimate of the relaxation time
\begin{equation}
\tau_{r}^{d}=  \frac{d^2 }{ D} \sim 80~{\rm s}.
\end{equation}

 

\begin{thebibliography}{99}
\bibitem{comment:CoarseGrain} The continuous order parameter field provides a coarse-grained description of the microscopic magnetic impurity system, which is valid when the spacing between impurities is small compared with the inverse ordering wavevector and at low temperatures where longitudinal magnetization fluctuations can be ignored.
\bibitem{Comment:OneBand} Note that in this case the electronic system becomes effectively spinless, and the $g_1$ and $g_2$ processes in the standard spinful Tomonaga-Luttinger model take the same bosonized form.
\bibitem{comment1} The cutoff $a$ parametrizes a finite bandwidth. In numerical calculations, we use $a= 1/k_F^{'}$ where $k_F^{'}$ is the Fermi vector in the original (non-rotated) reference frame.
\bibitem{commentM1} Time-ordered quantities are indicated with tildes.
\bibitem{comment2b} For the cut-off frequency $\omega_0 = 2\pi u_{\rm eft}/ L$ to be considered as an infinitesimal quantity, we need $\omega_0\tau_{s} <<  1$, where $\tau_{s}\sim a/u_{\rm eff}$ is the correlation time of the spin susceptibility. Using $a\sim 1/ k_F^{'}$, this implies that the system length should be much larger than the Fermi wavelength, $L>>\lambda_F$.
\bibitem{CommentM2}Note that a recent first-principle study of Permalloy reports signatures of nonlocality in the damping~\cite{Yuan:arxiv}.
\bibitem{Comment_kf} For RKKY systems, the wave vector $k_F$ in the rotated frame is twice the lab-frame wave vector $k_F^{'}$, i.e., $k_F= 2 k_F^{'}$.
\end{thebibliography}
\end{document}